\documentstyle[12pt]{article}
\newcommand{\scrip}{{\cal I}^+}
\newcommand{\scrim}{{\cal I}^-}
\newcommand{\be}{\begin{equation}}
\newcommand{\ee}{\end{equation}}
\newcommand{\bal}{\begin{array}{l}}
\newcommand{\eal}{\end{array}}
\newcommand{\pat}{\partial}
\newcommand{\bra}{\langle}
\newcommand{\ket}{\rangle}
\newcommand{\lam}{\lambda}
\newcommand{\laminv}{\frac{1}{\lambda}}
\newcommand{\kap}{\kappa}
\newcommand{\yp}{y^+}
\newcommand{\ym}{y^-}
\newcommand{\sigp}{\sigma^+}
\newcommand{\sigm}{\sigma^-}
\newcommand{\Om}{\Omega}
\newcommand{\om}{\omega}
\newcommand{\bea}{\begin{eqnarray}}
\newcommand{\eea}{\end{eqnarray}}
\newcommand{\lessapp}{\stackrel{<}{\sim}}
\newcommand{\greapp}{\stackrel{>}{\sim}}

\begin{document}
\begin{titlepage}

\begin{flushright}
MIT-CTP-2272 \\
hep-th/9312194
\end{flushright}

\vskip 0.9truecm

\begin{center}
{\large {\bf EVAPORATING BLACK HOLES AND ENTROPY$^*$ \\ }}
\end{center}

\vskip 1.5cm

\begin{center}
{\sc Esko Keski-Vakkuri} and {\sc Samir D. Mathur} \\
\vskip 0.4cm
{\it Center for Theoretical Physics \\
 Laboratory for Nuclear Science \\
 and Department of Physics \\
 Massachusetts Institute of Technology \\
 77 Massachusetts Avenue, Cambridge MA 02139, USA }
\end{center}

\vskip 1.2cm

\begin{center}
{\large {\rm December 27th, 1993} \\ }
\end{center}

\vskip 0.7cm

\begin{center}
(Submitted to Phys. Rev. D15)
\end{center}

\vskip 1.3cm

\begin{center}
{\small {\bf Abstract}}
\end{center}

\rm

\noindent
We study the Hawking radiation for the geometry of an evaporating
1+1 dimensional black hole. We compute Bogoliubov coefficients
and the stress tensor. We use a recent result of Srednicki to estimate
the
entropy of entanglement produced in the evaporation process, for the
1+1 dimensional hole and for the 3+1 dimensional hole. It is found that
the one space dimensional result of Srednicki is the pertinent one to
use, in both cases.

\vskip 2.5cm

\begin{flushleft}
$^*$ {\small
This work was supported in part by funds provided by the U.S. Department
of Energy (D.O.E.) under contract $\sharp$DE-AC02-76ER03069.}
\end{flushleft}

\end{titlepage}

\vfill\eject

\baselineskip 0.65 cm

{\large {\bf 1. Introduction}}

\bigskip

Several interesting phenomena are related to the discovery of Hawking
radiation \cite{H}.  It is intriguing that
black holes seem to obey laws of thermodynamics
\cite{BEK}.  The information contained in the matter which made
up the black hole is lost into the singularity. Hawking radiation
appears in the evaporation of the hole, but the outgoing modes are
not in a pure
state; instead they are mixed with modes of the field that fall into
the singularity. The precise significance of black hole thermodynamics,
and its relation to the ordinary ideas of thermodynamics and
information theory, are matters of debate.

Recently the discovery of  1+1 dimensional models for black holes
\cite{CGHS,RST} has led to a more accurate understanding of the
semiclassical features of black hole geometry and Hawking  radiation.
In particular the model of \cite{RST} (the RST model) can be exactly
solved to yield the   semiclassical geometry of a black hole formed
by a shock wave of infalling matter, and evaporating by massless
scalars to an endpoint (the `thunderpop').  It may even be possible
to obtain a complete quantum gravity plus matter description of the
black hole evaporation process
in 1+1 dimensions \cite{SVV}.

In this paper we  study  some features of the semiclassical geometry
and Hawking radiation in semiclassical  models. For the RST model of
the evaporating hole we compute the Bogoliubov coefficients. We
perform a point splitting calculation to compute the stress tensor at
$\scrip$. We also compute the stress tensor in
the evaporating region using the anomalous trace of the matter stress
tensor. The RST model is solved also for the case where the hole is
formed by one shock wave,  evaporation occurs for a time, and then a
second shock wave increases the mass of the hole again.  (This
geometry is used for clarifying some aspects of the entropy produced
by the hole, as discussed below.)

We then study the `entropy of entanglement' of the Hawking radiation,
by two methods. We can compute the density matrix obtained by
tracing
the field modes inside the horizon. This was the approach taken by
Hawking and also carried out in \cite{GN} for the 1+1 dimensional
case, and it yields a density matrix that is close to  thermal after
the initial stage of the collapse and formation of the hole.
We consider the case where the hole has a finite lifetime (due to the
evaporation) and thereby estimate the entropy of the entire radiation
produced.  We then compare this result to that obtained by using a
calculation of Srednicki \cite{SRED}.
Srednicki considered a scalar field in flat Minkowski space
in the vacuum state, and 'traced out' the degrees of freedom
inside a ball of radius R. The entropy of the
reduced density matrix is the 'entropy of entanglement' between
the region inside the ball  and its complement.
The entropy S depends on R and also on the ultraviolet cutoff,
which gives the 'sharpness of separation' between the regions.
In the one
space-dimension case, the infrared cutoff also appears. We find that
both for the 1+1 dimensional black hole and for the 3+1 dimensional
hole the one space dimension result of Srednicki is the pertinent one
to use, and the leading dependence of the entropy on the black hole
mass is reproduced.

The plan of this paper is as follows. In section 2 the RST model is
reviewed. In section 3 the Bogoliubov coefficients
for a scalar field in the evaporating black hole background are
computed. Section 4 contains the calculation of stress-tensor.
Section 5 studies the two
shock wave solution.
We discuss the entropy of the Hawking
radiation for 1+1 dimensional black holes in section 6,
and for 3+1 dimensional black holes in section 7.
Finally, a discussion is presented in section 8.

\bigskip

{\large {\bf 2. The RST model}}

\bigskip

The model of Russo, Susskind and Thorlacius (RST) \cite{RST} is a
modified
version of the model of
two dimensional dilaton gravity coupled to quantum matter
introduced in \cite{CGHS}. The key idea of RST was to introduce an
additional counterterm which restores a global symmetry originally
present
in the classical dilaton gravity + matter action. This allowed them to
solve the theory analytically in the large $N$ limit. The properties
of the RST model have been extensively discussed in the
literature \cite{RST,S,ST} so we will just mention the facts
we will need for later use.

The semiclassical effective action of RST can be written as
follows
\bea
  S &=& \frac{1}{2\pi} \int d^2x \sqrt{-g} \{ (e^{-2\phi} -\frac{N}{24} \phi)
  R + 4e^{-2\phi} [(\nabla \phi)^2 + \lam^2] - \frac{1}{2}
   \sum^N_{i=1} (\nabla f_i)^2 \} \\ \nonumber
   \mbox{} &-& \frac{N}{96} \int d^2x \sqrt{-g(x)} \int d^2x' \sqrt{-g(x')}
    R(x) G(x,x') R(x') \ ,
\eea
where $R$ is the 1+1 dimensional scalar curvature, $\phi$ is the
dilaton field and $f_i,\ i=1\ldots ,N$ are $N$ massless conformal
scalar fields. $G(x,x')$ is the Green's function for the
d'Alembertian in curved space. The constant $\lam $ plays the
role of Planck mass.

The analysis of the semiclassical equations of motion that
follow from the action (1) can be simplified by the following two
steps. First, one can write the metric in the conformal gauge, given
by
$$
  g_{\pm \mp} = -\frac{1}{2} e^{2\rho}\ , \ g_{\pm \pm} = 0 \ ,
$$
where $x^{\pm} = x^0 \pm x^1$. Secondly, one can
make a field redefinition
and introduce the fields
\bea
  \Om &=& \kap^{-1} e^{-2\phi} + \frac{\phi}{2}
         + \frac{1}{4} \ln \frac{\kap}{4} \\ \nonumber
  \chi &=& \kap^{-1} e^{-2\phi} + \rho -\frac{\phi}{2} -
             \frac{1}{4} \ln (4\kap ) \ ,
\eea
where $\kap \equiv \frac{N}{12}$.
The coordinates $x^{\pm}$ can be fixed
so that $\Om = \chi$, so the dilaton field $\phi$ differs from the
conformal
factor of the metric $\rho$ by a constant.
The matter fields are assumed to reflect from the strong coupling
boundary $\Om = \Om_{cr} =\frac{1}{4}$ \cite{RST}.
The semiclassical equations can now be reduced to
\bea
    \pat_+ \pat_- f_i &=& 0 \\ \nonumber
    \pat_+ \pat_- \Om &=& -\lam^2 \\ \nonumber
    \kap \pat^2_{\pm} \Om &=& -\pi T^f_{\pm \pm}
                          + \kap t_{\pm} (x^{\pm}) \ .
\eea
Here $T^f_{\pm \pm}$ are the components for outgoing and ingoing
matter energy of the stress-tensor, which is defined as follows.
Since the classical
matter action is written as
\be
 S_f = -\frac{1}{4\pi} \int d^2 x \sqrt{-g} \sum^N_{i=1}
   (\nabla f_i)^2 \ ,
\label{Smat}
\ee
the stress tensor is
$$T^f_{\mu \nu } = \frac{-2}{\sqrt{-g}}
 \frac{\delta S_f}{\delta g^{\mu \nu}} \ .$$
The components representing outgoing and ingoing matter
are normalized as
$$T^f_{\pm \pm} = \frac{1}{2\pi}
  \sum_i \pat_{\pm} f_i \pat_{\pm} f_i \  ,$$
in the conformal gauge.

The functions $t_{\pm}(x^{\pm})$ are fixed
by boundary conditions. We assume that the
incoming matter energy flux at $\scrim$
vanishes sufficiently rapidly at early and late times.
Then,
$t_+(x^+) = 1/4(x^+)^2$. The solution for the field $\Om$ can now be
found to be
\be
 \Om = -\lam^2 x^+(x^- +\frac{\pi}{\kap \lam^2} P_+(x^+))
    + \frac{\pi}{\kap \lam} M(x^+) - \frac{1}{4}\ln (-4\lam^2 x^+ x^-) \ ,
\ee
where
\bea
 M(x^+) &=& \lam \int^{x^+}_0 du \ u
         \ T^f_{++}(u) \\ \nonumber
 P_+(x^+) &=& \int^{x^+}_0 du \ T^f_{++}(u) \ .
\label{eq:mpp}
\eea
Consider now an incoming matter shock wave that
carries energy $M$.
The stress tensor is then given by
\be
T^f_{++}(x^+) = \frac{M}{\lam x^+_0} \delta (x^+ - x^+_0) \ ,
\label{shock1}
\ee
which is the only non-vanishing component. We substitute this
in the equations (6) above. Following \cite{S}, we
set $\lam x^+_0 = 1$.

After solving the RST equations one finds the following results.
The Penrose diagram for the evaporating
black hole spacetime is given in Fig. 1.

The spacetime is seen to be divided into three regions.
The 'lowest' region I, before the
incoming shockwave at $x^+_0$, is the linear dilaton vacuum,
bounded from the left by the timelike
strong coupling boundary
$\Om=\Om_{cr}$. The incoming shockwave forms the
black hole by causing the boundary $\Om =\Om_{cr}$
to become spacelike. It can be shown that the scalar
curvature $R$ diverges at the spacelike portion
of the boundary. The apparent horizon also forms
after the incoming shockwave. Following \cite{RST}, the
apparent horizon is defined by the
condition $\pat_+ \phi =0$. After the black hole
forms, it starts to evaporate, and the apparent
horizon shrinks until it meets the singularity
at the endpoint of evaporation. At the endpoint
of evaporation a short (delta function) burst
of negative energy is seen to emerge from the black hole.
This is called the 'thunderpop' \cite{RST}, and it travels along
the null line $x^- =x^-_s$ to future null infinity.
($x^-_s$ is a light-cone coordinate of the endpoint
to be specified later.) Thus the region II bounded
by the thunderpop and the incoming shockwave is
the curved region of the evaporating black hole.
After the thunderpop, the
spacetime becomes again flat and the
boundary $\Om =\Om_{cr}$ becomes timelike.
The corresponding
region III is a linear dilaton vacuum.

In the linear dilaton vacuum region I the metric is
$$ds^2 = (\lam^2 x^+ x^-)^{-1} dx^+ dx^- \ .$$
We can write it as
$ds^2 = -d\yp d\ym$ using
coordinates
\bea
\ym &=& -\frac{1}{\lam} \ln (-\lam x^-) \\ \nonumber
\yp &=& \frac{1}{\lam} \ln (\lam x^+) - \yp_0 \ .
\eea
The shift $\yp_0$ is introduced
to set the origin of the coordinate $\yp$ to the point $A$ where
the reflected ray of the event horizon (see Fig. 1.) meets $\scrim$.

The event horizon, the singularity and the apparent horizon
meet at point E, the end point of evaporation.
In our conventions, its coordinates are
\be
  (x^-_s,\ x^+_s) = ( -\frac{\pi M}{\kap \lam^2}
  (1 - e^{-\frac{4\pi M}{\kap \lam}})^{-1},\ \frac{\kap }{4\pi M}
  (e^{\frac{4\pi M}{\kap \lam}} -1) ) \ .
\ee
We can now specify what the shift $\yp_0$ is. In region I the reflecting
boundary $\Om = \Om_{cr} = \frac{1}{4}$ is the line
$$\yp = \ym + \frac{1}{\lam} \ln \frac{1}{4} - \yp_0 \ . $$
Reflecting the line
$x^- = x^-_s$ (off the boundary $\Om=\Om_{cr}$)
back to $\scrim$, we find that the point A has
$\yp_A = -\laminv \ln (-\lam x^-_s) + \laminv \ln \frac{1}{4} - \yp_0$.
Setting $\yp_A=0$ yields $\yp_0=-\laminv \ln (-4\lam x^-_s)$.

The region II is the curved region of the evaporating black hole. It
can be joined continuously but not smoothly with region III along the
line $x^- = x^-_s$, with
region III being again a linear dilaton vacuum, but with
the coordinate $x^-$
shifted. From the joining conditions the metric in III
can be found to be
$$ds^2 = (\lam^2 x^+(x^-+\frac{\pi M}{\lam^2 \kap}))^{-1}
          dx^+ dx^- \ .  $$
If one makes the coordinate transformation
\bea
 \sigp &=& \laminv \ln (\lam x^+) \\ \nonumber
 \sigm &=& -\laminv \ln (\frac{\lam x^- + \frac{\pi M}{\lam \kap}}
                        {\lam x^-_s + \frac{\pi M}{\lam \kap}}) \ ,
\eea
this metric becomes $ds^2 = -d\sigp d\sigm$.
The coordinate $\sigm$ has been normalized so
that the thunderpop is
at $\sigm =0$. The reflecting boundary $\Om =\Om_{cr}$
in region III is at
$\sigp = \sigm + \laminv \ln (\lam x^+_s)$.

The metric in II becomes asymptotically flat near
$\scrip$. We can extend the coordinates $\sigma^{\pm}$ from region III
into region II in the neighbourhood of $\scrip$.  Then the metric
in region II also has the asymptotic form
$ds^2 \rightarrow -d\sigp d\sigm$ near $\scrip$.
Thus, $\sigma^{\pm}$ are the physical coordinates near $\scrip$.

Finally, we identify some points of interest in the Penrose diagram.
The point A where the reflection of the null line $x^-=x^-_s$
meets $\scrim$ we already set to be at $\yp_A=0$. The point B is the
projection of the end point E along a null ray
to past null infinity. We find
it to be located at $\yp_B=\frac{4\pi M}{\kap \lam^2}$. The point C is
defined
by projecting the point where the apparent horizon meets the incoming
shockwave along a null ray to future null infinity.
It is at
$\sigm_C = -\laminv \ln
(\frac{\kap \lam}{4\pi M} (\exp (\frac{4\pi M}{\lam \kap}) - 1))$. Since
$M >> \kap \lam$, to a good accuracy
$\sigm_C = -\frac{4\pi M}{\lam^2 \kap}$.
The absolute value of $\sigm_C$
is thus the total (retarded) time of evaporation
of the black hole.

\bigskip

{\large {\bf 3. Bogoliubov transformations}}

\bigskip

In this section we calculate the Bogoliubov coefficients for the relation
between the natural mode functions in the 'in' region close to $\scrim$
and the 'out' region close to $\scrip$.
In \cite{H} the Bogoliubov coefficients were estimated for modes
travelling close to the horizon that forms in a spherical collapse
of a star.
In \cite{GN} the Bogoliubov coefficients were computed for the 1+1
dimensional eternal black hole geometry, {\em i.e.} without taking into
account the backreaction on the metric due to the Hawking
evaporation.
Our notations follow those in \cite{GN} where
we refer to for all introductory steps.

We choose the following form for modes at $\scrim ,\ \scrip$
respectively
\bea
   u_{\om} &=& \frac{1}{\sqrt{2\om}}
                e^{-i\om \yp} \ \ \ \ \ (in) \\  \nonumber
   v_{\om} &=& \frac{1}{\sqrt{2\om}}
                e^{-i\om \sigm} \ \ \ \ \ (out) \ ,
\eea
where the normalization factor is in agreement with
the form (\ref{Smat}) of the
matter action.

We define the Bogoliubov coefficients $\alpha_{\om \om'}$
and $\beta_{\om \om'}$  for the relation between the 'in' and 'out'
modes as follows
\be
  v_{\om} = \int^{\infty}_{0} d\om' \ [\alpha_{\om \om'} u_{\om'}
             + \beta_{\om \om'} u^*_{\om'} ] \ .
\ee
To compute the Bogoliubov coefficients, we pull
the 'out' mode back to $\scrim$. First we divide
the mode into two parts at the point $\sigm =0$.
The 'upper' piece
$v_{\om} = \frac{1}{\sqrt{2\om}} e^{-i\om \sigm} \theta (\sigm )$
reflects from the timelike boundary in region III
without experiencing any blueshift.
At $\scrim $ it becomes
$$
  v_{\om} = \frac{1}{\sqrt{2\om}} e^{-i\om (\yp -\yp_B )}
          \theta (\yp -\yp_B ) \ ,
$$
where $\yp_B$ was defined in the previous section.

However, the 'lower' piece
$v_{\om} = \frac{1}{\sqrt{2\om}} e^{-i\om \sigm} \theta (-\sigm )$ gets
distorted. It first experiences a blueshift when pulled back to region I.
This is done by replacing the coordinate $\sigm $ with the coordinate
$\ym $ using the relation
$$
 \sigm = -\laminv \ln [\frac{ -e^{-\lam \ym} + \pi M/\lam \kap}
         {\lam x^-_s + \pi M/\lam \kap}] \ .
$$
Then we reflect the mode from the timelike boundary
in region I back to $\scrim $, by
replacing $\ym $ with
$$
 \ym = \yp + \yp_0 -\laminv \ln  \frac{1}{4}
      = \yp -\laminv \ln (-\lam x^-_s) \ .
$$
The final relation between the
coordinates $\sigm$ and $\yp$ can then
be written as
\bea \label{eq:rel1}
  \sigm &=& -\laminv \ln [\frac{ \lam x^-_s e^{-\lam \yp}
                       + \pi M/\lam \kap}
           {\lam x^-_s + \pi M/\lam \kap}] \\ \nonumber
        &=& -\laminv \ln [\lam \Delta (e^{-\lam \yp} - C)] \ ,
\eea
where
\bea
 \lam \Delta &\equiv & \frac{\lam x^-_s}
                     {\lam x^-_s +\pi M/\lam \kap}
              =  e^{4\pi M/\lam \kap} \\ \nonumber
  C &\equiv & -\frac{\pi M/\lam \kap}{\lam x^-_s}
                = 1 - e^{-4\pi M/\lam \kap} \ .
\eea

As an important aside, notice that (\ref{eq:rel1})
implies a relation between a small distance $d\sigm$ at $\scrip$
centered at $\sigm$ and the corresponding small distance $d\yp$ at
$\scrim$ which results from mapping the former distance
along null rays which reflect from
the boundary back to the past null infinity.
This relation can be found to be
\be
 d\yp = \{ 1+(e^{4\pi M/\kap \lam}-1)
            e^{\lam \sigm} \}^{-1} d\sigm \ .
  \label{eq:squeeze1}
\ee
The result (\ref{eq:squeeze1}) tells us
that a distance $d\sigm$ at $\scrip$
(before the endpoint of evaporation)
maps to a much smaller distance
$d\yp$ at $\scrim$, with the 'squeeze factor' becoming
exponentially larger as
the black hole evaporates.
This observation will turn out to be crucial in applying
the Srednicki calculation for black holes, as will be
discussed in the section 6.

Returning back to the
behaviour of the modes, (\ref{eq:rel1}) tells
us that the 'lower' part of $v_{\om}$ becomes
\be
 v_{\om} = \frac{1}{\sqrt{2\om }}
    \exp \{ \frac{i\om}{\lam} \ln
    [\lam \Delta (e^{-\lam \yp } - C)] \}
     \theta (-\yp )
\ee
when pulled back to $\scrim $.

It is now straightforward to proceed to find
the Bogoliubov coefficients.
The result is
\bea \label{bogol}
 \alpha_{\om \om '} &=&
     \frac{1}{2\pi} \sqrt{\frac{\om'}{\om}}
               \{ \int^0_{-\infty} d\yp
 \exp \{ \frac{i\om}{\lam}
  \ln [\lam \Delta (e^{-\lam \yp}-C)] + i\om' \yp \} \\ \nonumber
        & & \ \ \ \mbox{} + e^{i\om ' \yp_B} \int^{\infty}_0 dz^+
         \exp \{ -i(\om -\om' ) z^+ \} \}  \\ \nonumber
 \beta_{\om \om '} &=&
      \frac{1}{2\pi} \sqrt{\frac{\om'}{\om}}
               \{ \int^0_{-\infty} d\yp
    \exp \{ \frac{i\om}{\lam}
  \ln [\lam \Delta (e^{-\lam \yp}-C)] - i\om' \yp \} \\ \nonumber
         & & \ \ \ \mbox{} + e^{-i\om ' \yp_B} \int^{\infty}_0 dz^+
         \exp \{ -i(\om +\om' ) z^+ \} \} \ .
\eea
These results resemble the ones of Giddings and Nelson (GN), with
the following
three differences: 1) there are additional terms
resulting from the 'upper'
part of the mode $v_{\om}$, {\em i.e.}, the part
after the endpoint of evaporation, 2) $\lam \Delta$
is different, and 3) we have
$C = 1-e^{-4\pi M/\kap \lam}$ while GN had $C=1$.
(Of course, $C \approx 1$ since
$M/\kap \lam >> 1$. However, the small difference turns out to be
significant if one tries to investigate
the behaviour of the modes very near the endpoint.)

The integrals in the above expressions can be evaluated further.
Substituting first $x=e^{\lam \yp}$ and then $t=Cx$, we get
\bea
\alpha_{\om \om'} &=&
\frac{1}{2\pi}\  (\frac{\om'}{\om})^{1/2} \ \{
  \ (\lam \Delta )^{i\om /\lam}
  \ C^{i(\om -\om' )/\lam}  \ \int^C_0 dt \ (1-t)^{i\om /\lam }
  \ t^{-1-i(\om -\om' )/\lam } \\ \nonumber
   & & \ \ \ \ \mbox{} + e^{i\om' \yp_B}
   \ \frac{-i}{\om -\om' -i\epsilon} \} \ .
\eea
The remaining integral can be identified as an Incomplete Beta
function.
The coefficient $\beta_{\om \om'}$ is computed similarily. The final
expressions are
\bea
 \alpha_{\om \om'} &=& \frac{1}{2\pi \lam}
  \ (\frac{\om'}{\om })^{1/2}\ \{ (\lam \Delta)^{i\om /\lam}
  \ C^{i(\om +\om')/\lam}
 \ B_C (-\frac{i\om}{\lam}
     +\frac {i\om'}{\lam} +\epsilon, i+\frac{i\om}{\lam}) \\ \nonumber
   & & \ \ \mbox{} - i\lam
     \ e^{i\om' \yp_B} \ (\om -\om' -i\epsilon')^{-1} \} \\ \nonumber
 \beta_{\om \om'} &=& \frac{1}{2\pi \lam}
  \ (\frac{\om'}{\om })^{1/2}\ \{ (\lam \Delta)^{i\om /\lam}
  \ C^{i(\om -\om')/\lam}
 \ B_C (-\frac{i\om}{\lam}
      -\frac{i\om'}{\lam} +\epsilon, i+\frac{i\om}{\lam}) \\ \nonumber
   & & \ \ \mbox{} - i\lam
      \ e^{-i\om' \yp_B} \ (\om +\om' -i\epsilon')^{-1} \} \ .
\eea
Note that in the semiclassical approximation the thunderpop
is a delta function at $\scrip$, thus there is a part of the field
modes that is not captured by the modes $e^{-i\om \sigm}$ at $\scrip$
for any finite range of $\om$. Thus the Bogoliubov coefficients
computed
below need to be supplemented with an
infinite frequency component to
completely describe the field at $\scrip$.

Let us now turn
to the issue of examining the nature of
the radiation by studying the
approximate behaviour of the Bogoliubov coefficients. We expect
outgoing thermal
radiation at constant temperature $\frac{\lam}{2\pi}$ to be
seen in the
region $\sigm \in ( -\frac{4\pi M}{\kap \lam^2} , 0 )$ of
$\scrip $, except
perhaps at the beginning and very end of the evaporation process.
The coordinate $\yp$ corresponding to this region is
very small, so we
can approximate
\be
  \ln [ \lam \Delta (e^{-\lam \yp} -C)] \approx
  \ln [1-e^{4\pi M/\kap \lam} \lam \yp ] \ .
\label{eq:approx1}
\ee
If we are not too close to the
endpoint, the term 1 in (\ref{eq:approx1})
is negligible. For the essence of
Hawking radiation, we have the frequency
$\om \sim \lam$ at $\scrip$ and very high frequencies
$\om' \sim \lam e^{4\pi (M+\yp)/\kap \lam}$ at $\scrim$.
For such $\om ,\om' $ we can ignore the second integrals in the
expression (\ref{bogol}) for
$\alpha_{\om \om'} ,\beta_{\om \om'}$. We
therefore get
\bea
 \alpha_{\om \om '} &\approx &
     \frac{1}{2\pi} \sqrt{\frac{\om'}{\om}}
               \{ \int^0_{-\infty} d\yp
 \exp \{ \frac{i\om}{\lam}
 \ln [ -e^{4\pi M/\lam \kap} \lam \yp] + i\om' \yp \} \} \\ \nonumber
   \beta_{\om \om '} &\approx &
      \frac{1}{2\pi} \sqrt{\frac{\om'}{\om}}
               \{ \int^0_{-\infty} d\yp
    \exp \{ \frac{i\om}{\lam}
  \ln [ -e^{4\pi M/\lam \kap} \lam \yp] - i\om' \yp \} \ .
\eea
This means that we get the thermal relation (see \cite{GN})
\be
   \alpha_{\om \om'} \approx -e^{\pi \om /\lam} \beta_{\om \om'}
\ee
and we see that the outgoing radiation is thermal with
constant temperature $T_H = \frac{\lam}{2\pi}$ in the region
$\sigm \in ( -\frac{4\pi M}{\kap \lam^2} , 0 )$ at $\scrip $.

\bigskip

{\large {\bf 4. The stress tensor}}

\bigskip

Since we have worked out the relations between the 'in' and 'out'
modes,
we can easily calculate the VEV of
the stress-energy at $\scrip$, {\em i.e.}
$\bra 0,in \mid T_{\mu \nu}(\sigm ) \mid 0,in\ket^{ren}$ by using the
point-splitting method. The non-vanishing
component of $\bra T_{\mu \nu} \ket^{ren}$
is the $--$ component.
Since we are computing the 'in' VEV at $\scrip$, we need
first the form
of the 'in' mode at $\scrip$. For $\sigm \leq 0$ it is
\be
 u_{\om} = \frac{1}{\sqrt{2\om}} \exp \{ \frac{i\om}{\lam}
 \ln [ \frac{1}{\lam \Delta} e^{-\lam \sigm}+C]\} \theta (-\sigm ) \ .
\ee
Again, for $\sigm >0$ (after the endpoint) there is no redshift
and it is then trivial to see that
the stress-tensor vanishes in this region. We concentrate only in the
region before the endpoint.
Using the point-splitting method, we first calculate
\be
 \bra T_{--}(\sigm )\ket = \lim_{\sigm_1, \sigm_2 \rightarrow \sigm}
  \frac{1}{2\pi} \frac{\pat}{\pat \sigm_1} \frac{\pat}{\pat \sigm_2}
 \int^{\infty}_{0} d\om \ u_{\om} (\sigm_1 )u^*_{\om}(\sigm_2 ) \ ,
\ee
where the LHS means the VEV with respect to the 'in'-vacuum. (We
use this notation in the following.)  For
$\sigm <0$ the step functions are just constant and can be ignored.
Taking the partial derivatives and making a series expansion
in $(\sigm_1 - \sigm_2)$ yields then a term which diverges
quadratically
in the limit, and a finite term. The divergent term
must be subtracted (it
is just the usual vacuum divergence) and the
finite term gives the
renormalized value for the stress-energy:
\be
 \bra T_{--} (\sigm ) \ket^{ren} = \frac{\lam^2}{48\pi}
 [ 1-\frac{1}{(1+(e^{4\pi M/\kap \lam}-1)
      e^{\lam \sigm})^2} ] \ .
\ee

This result is for one conformal scalar field, for
$N$ fields it must be multiplied by $N$. Note that in the region
$\sigm \in ( -\frac{4\pi M}{\kap \lam^2} , 0 )$ we can approximate
\be
  \bra T_{--} (\sigm ) \ket^{ren} \approx \frac{\lam^2}{48\pi}
        = \frac{\pi}{12} (T_H)^2 \ ,
\ee
which is the correct value for outgoing thermal radiation at temperature
$T_H=\frac{\lam}{2\pi}$.

(Comment: if one would use the {\em approximate}
behaviour
of the mode very near the end point, one would find that
$\bra T_{--} \ket^{ren} \rightarrow 0$ at roughly Planck distance from
the end point. One should not, however, trust such a local treatment
when dealing with modes.)

The above formula gave $\bra T_{--} \ket$
only at $\scrip$. However,
in 1+1 dimensions it is easy
to calculate $\bra T_{\mu \nu} \ket^{ren}$ {\em everywhere} by using
the trace anomaly, integrating the covariant conservation equation and
applying the boundary conditions (see \cite{HS} for details). Using
this route, we end up with the following results for the stress tensor
everywhere in region II:
\bea
  \bra T_{--} \ket^{ren} &=& \frac{N}{48\pi} \frac{1}{(x^-)^2}
  (\frac{e^{-2\rho}+1/4}{e^{-2\rho}-1/4})
  \{ (\frac{\lam^2 x^+ x^- + 1/4}
    {e^{-2\rho} -1/4})^2 -1 \} \\ \nonumber
  \bra T_{++} \ket^{ren} &=& \frac{N}{48\pi} \frac{1}{(x^+)^2}
    (\frac{e^{-2\rho}+1/4}{e^{-2\rho}-1/4})
 \{ (\frac{\lam^2 x^+ (x^- +\pi M/\kap \lam^2)+1/4}
    {e^{-2\rho} -1/4})^2 -1 \} \\ \nonumber
   \bra T_{+-} \ket^{ren} &=& -\frac{N\lam^2}{24\pi}
    \frac{1}{e^{-2\rho}-1/4} \{ 1 + \frac{e^{-2\rho}}
    {(e^{-2\rho}-1/4)^2}
  \frac{[\lam^2 x^+ (x^- +\frac{\pi M}{\kap \lam^2})+1/4]
   [\lam^2 x^+ x^- +1/4]}{\lam^2 x^+ x^-} \} \ ,
\eea
where $\rho$ is given implicitely via the equation
\be
   e^{-2\rho} + \frac{\rho}{2} = -\lam^2
    x^+(x^- + \frac{\pi M}{\kap \lam^2}) - \frac{1}{4}
   \ln [-\lam^2 x^+ x^- ] + \frac{\pi M}{\kap \lam} \ .
\ee
In the above the stress tensor is given in
the 'Kruskal' coordinates $x^{\pm}$ since
one is generally
interested in its behaviour in both sides of the apparent horizon.

\bigskip

{\large {\bf 5. Two incoming shock waves}}

\bigskip

In this section we present the solution of the RST equations in the
case of two incoming shock waves. We use this geometry later in
section 6
where we discuss the entropy of the black hole. We want
to first form a black hole of mass $M_0$, which then starts to
evaporate.
Then the second shock wave at a later time
carries additional energy to
the black hole. For instance, it could restore
the black hole back to its
original mass. The question then is: How much is the entanglement
entropy after the evaporation process?
If the second shock wave carries just enough energy to restore
the black hole back to its original mass, but not
more, is this entanglement
entropy related to Bekenstein
entropy? We discuss this question in section 6, here we will
just derive the results that we will use.

Recalling how we defined the incoming shock wave, it is clear that for
two incoming shock waves we should replace (\ref{shock1}) with
\be
   T^f_{++}(x^+) = \frac{M_0}{\lam x^+_0}
  \delta (x^+ - x^+_0) + \frac{M_1}{\lam x^+_1}
  \delta (x^+ - x^+_1) \ .
\ee
The first term on the RHS is the first
shock wave at $x^+_0$ (we set again
$\lam x^+_0 =1$) carrying energy $M_0$ and
the second term is a later
shock wave at $x^+_1$ carrying energy $M_1$. We do not
specify  $x^+_1,\ M_1$ here, but $x^+_1$ should be
chosen so that the black hole formed by
the first shock wave has not yet
evaporated away when the second shock wave
reaches it. On could think of $M_1$ being the energy needed to
restore the black hole back to its
original size, but for now that is not essential.

The spacetime curve for the apparent horizon can be found by solving
the equation $\pat_+\Om =0$ (this follows from setting
$\pat_+\phi =0$ for the
dilaton field; which is the RST definition of the apparent
horizon). The solution is
\be
 x^+ = -\frac{1}{4\lam}
  \frac{1}{(x^- + \frac{\pi M_0}{\kap \lam^2}
      +\frac{\pi M_1}{\kap \lam^3 x^+_1} \theta(x^+-x^+_1))} \ ,
\ee
for $x^+>x^+_0$. The boundary of the spacetime is the critical line
$\Om =\Om_{cr}=\frac{1}{4}$, which defines another curve
$x^+=x^+(x^-)$.
The (final) endpoint of evaporation is where the above two curves
meet.
We find it to be located at
\bea \label{E2}
 x^-_s &=& -\frac{\pi}{\kap \lam^2} (M_0+\frac{M_1}{\lam x^+_1})
     (1-e^{-4\pi (M_0+M_1)/\kap \lam})^{-1} \\ \nonumber
 x^+_s &=& \frac{\kap}{4\pi (M_0+\frac{M_1}{\lam x^+_1})}
     (e^{4\pi (M_0+M_1)/\kap \lam} -1) \ .
\eea

We can now join the region II with the linear dilaton vacuum
in region III with an appropriate shift of the coordinate
$x^-$. For the metric in region III, we find
$$
  ds^2 = \frac{dx^+dx^-}
 {\lam^2 x^+(x^- + \frac{\pi}{\kap \lam^2}
        (M_0+M_1/\lam x^+_1))} \ .
$$
This in turn tells us how to define the coordinates $\sigma^{\pm}$
which are the 'physical' coordinates near $\scrip$.
We define
\bea
 \sigp &=& \laminv \ln (\lam x^+) \\ \nonumber
 \sigm &=& -\laminv \ln [\frac{\lam x^- +\frac{\pi}{\kap \lam}
            (M_0+M_1/\lam x^+_1)}
    {\lam x^-_s +\frac{\pi}{\kap \lam}(M_0+M_1/\lam x^+_1)}] \ ,
\eea
On the other hand, we still have the 'physical'
coordinates in region I
\bea
 \yp &=& \laminv \ln (\lam x^+) - \yp_0 \\ \nonumber
 \ym &=& -\laminv \ln (-\lam x^-) \ ,
\eea
where $\yp_0 =-\laminv \ln (-4\lam x^-_s)$, with
$x^-_s$ given now by (\ref{E2}).
The reflecting boundary in region I is the line
$\yp = \ym + \laminv \ln \frac{1}{4} - \yp_0$.

It is straightforward
to see that the relation between the coordinates $\sigm$, $\yp$
now becomes
\be
  \sigm = -\laminv \ln [\lam \Delta'(e^{-\lam \yp}-C')] \ ,
\ee
where
\bea
  \lam \Delta' &=& e^{4\pi (M_0+M_1)/\lam \kap} \\ \nonumber
  C' &=& 1-e^{-4\pi (M_0+M_1)/\lam \kap} \ .
\eea

As we noticed in section 3, this implies
a relation between a distance $d\sigm$ at $\scrip$,
centered at $\sigm$ and the corresponding small distance $d\yp$ at
$\scrim$ . In the two shock wave case, the relation is
\be
 d\yp = \{ 1+(e^{4\pi (M_0+M_1)/\kap \lam}-1)
   e^{\lam \sigm} \}^{-1} d\sigm \ .
\label{eq:squeeze2}
\ee
Again, a distance $d\sigm $in $\scrip$ (before the endpoint of
evaporation)
maps to a much smaller distance $d\yp$ in $\scrim$
and this 'squeeze factor'
becomes exponentially large
as the black hole evaporates.
In the present case the 'squeeze factor' reaches the
value $\sim e^{4\pi(M_0+M_1)/\kap \lam}$ which
exceeds the value $\sim e^{4\pi M_0/\kap \lam}$ that
would be obtained in the absence of the second
shock wave.
The result (\ref{eq:squeeze2}) will be used in the next section.

\bigskip

{\large {\bf 6. Entropy for 1+1 dimensional black holes}}

\bigskip

As argued by Hawking, the process of pair creation by the
gravitational field of the black hole creates a state which is
`mixed' between the regions exterior and interior to the horizon. If
we compute the reduced density matrix that describes the field in the
exterior region, then the entropy computed for this density matrix
gives the `entropy of entanglement'  \cite{PAGE} between the interior
and exterior regions of the hole.
If we do not take into account the backreaction from the created
radiation then an infinite number of particle pairs are produced and
the entropy of entanglement will be found to be infinite also. But in
the simple RST model we can estimate the produced entropy for the
semiclassical geometry that includes backreaction.  We shall do this
in two ways:
\begin{enumerate}
\item  We directly compute
\be S~=~-Tr\{\rho \ln\rho\} \ee
where $\rho$ is the reduced density matrix describing the field in
the exterior region. Here we use the fact that $\rho$ is to a good
approximation a thermal density matrix.
\item We use the result of Srednicki described in the introduction. Thus
a complete spacelike hypersurface is considered, and the part
corresponding to the interior of  the
horizon is assumed to be, effectively, the traced over region
considered in this approach.
\end{enumerate}

1. The essential idea is to define reasonably localised regions on
$\scrip$ such that the density matrix can be described as
approximately thermal in those regions. Hawking presented this
analysis for the 3+1 dimensional black hole \cite{H}; it was worked
out  explicitly for the 1+1 case in \cite{GN} (without
backreaction). We follow the notations
of \cite{GN} in the following. Define a complete set of orthonormal
wavepackets on $\scrip$:
\be v_{jn}~=~\epsilon^{-1/2}\int^{(j+1)\epsilon}_{j\epsilon}d\omega
\ e^{ 2\pi i \omega n/\epsilon}\ v_\omega \ ,
\label{packet}
\ee
where $j=0,1,2,\ldots $ and $n$ are integers.
The wavepacket $v_{jn}$ is peaked about
$\scrip$ coordinate $\sigm =2\pi n/\epsilon$, has a spatial width
$\sim \epsilon^{-1}$ and a frequency $\omega_j \approx j\epsilon$. In
this basis the reduced density matrix obtained by tracing out the
field inside the black hole is
\be
\rho~=~N^{-1} \sum_{\{ n_{jn}\} }
e^{-\frac{2\pi}{\lam}\sum_{jn}n_{jn}\omega_j}|\{ n_{jn}\} \ket
\bra \{ n_{jn}\} | \ .
\label{roo}
\ee
This has the form of a thermal density matrix. We can write
\be \rho~=~\prod_{jn}\rho_{jn} \ee
with
\be \rho_{jn}~=~N_{jn}^{-1}
 \sum_{n_{jn}=0}^{\infty} r(n_{jn})
|n_{jn} \ket \bra n_{jn}| \ee
\be r(n_{jn})~=~e^{-\frac{2\pi}{\lam} n_{jn} \om_j} \ . \ee
We get
\be S~=~\sum_{jn} S_{jn} \ee
where
\be
S_{jn}~=~-Tr_{(jn)}\{\rho_{jn}\ln
  \rho_{jn}\}~=~-\sum_{{n_{jn}}=0}^{\infty} r(n_{jn})
   \ln r(n_{jn}) \ . \ee
A brief computation gives
\be S_{jn}~=~\beta \omega_j (e^{\beta
\om_j}-1)^{-1}~-~\ln(1-e^{-\beta\om_j}) \ . \ee
Thus
\bea \label{Sn}
S_n~=~\sum_{j=0}^\infty S_{jn} &=& \sum_{j=0}^\infty \{ \beta
\omega_j (e^{\beta \om_j}-1)^{-1}~-~\ln
   (1-e^{-\beta\om_j}) \} \\ \nonumber
 &\rightarrow& \int_0^{\infty} dj \{ \beta j \epsilon
     (e^{\beta j
 \epsilon}-1)^{-1}~-~\ln
 (1-e^{-\beta j
  \epsilon}) \} ~=~{\pi^2\over 3\beta\epsilon} \ .
\eea
The entropy from $N$ evaporating fields is
\be
S=N\sum_n S_n  \ .
\label{SumSn}
\ee
The separation between wavepackets (\ref{packet})  is
$\Delta\sigma^-=2\pi/\epsilon$.
Thus in time $T$ the number $n$ ranges from $n=1$ to $n=\epsilon
T/2\pi$.
{}From (\ref{Sn}, \ref{SumSn}) we get
\be S={N\pi T\over 6\beta} \ . \ee
The total evaporation time is $T={48\pi M\over N\lambda^2}$. This
gives the estimate of the total entropy created in the Hawking
radiation:
\be S_{\rm total}~=~{4\pi M\over \lambda} \ .
\label{Stotal}
\ee
Note that this is {\em twice} the Bekenstein
entropy, which for the 1+1
dimensional hole is $S_{\rm Bek}=\frac{2\pi M}{\lambda}$.
The result (\ref{Stotal}) is the entropy of
a thermal distribution of bosons at
temperature $\beta^{-1}$ and with energy
$E=M$, which in one dimension is given by $S=2\beta E$. For a
discussion
of how the entropy of radiation at $\scrip$ relates to
the Bekenstein entropy, see \cite{Z}.

2. We now investigate the application of Srednicki's result to the
black hole. We split the discussion into 3 parts:
\begin{itemize}
\item{(i)} We recall the result of \cite{SRED}, and discuss
        the issue of infrared divergence.
\item{(ii)} We discuss how this result may be applied to
   the black hole, after making some plausible arguments for
  the required modifications.
\item{(iii)} We carry out the calculations for the entropy.
\end{itemize}

(i) The computation of Srednicki for the one space dimension case
may be described as follows.\footnote{Similar issues were
studied in \cite{B}.} We consider a free scalar field on a
1-dimensional lattice, with lattice
spacing $a$. Let this field be in the
vacuum state. We select a region of
length $R$ of this lattice and trace
over the field degrees of freedom outside this region. This gives a
reduced density matrix $\rho$, from which we compute $S= -Tr\{ \rho
\ln \rho \}$  which is the entropy of entanglement  of the selected
region
with the remainder of the lattice. It is
immaterial whether we trace over
the interior or exterior of the selected region; since the field was in a
pure state overall the entanglement entropy is the same in both cases.
This entropy is given by \cite{SRED}
\be S~=~ \kappa_1 \ln(R/a)~+~\kappa_2\ln(\mu R) \ ,
 \label{Ssred}
\ee
for one scalar field. (For $N$ species the result must
be multiplied by $N$.)
Here $\mu$ is an infrared cutoff.

The infrared term is
very sensitive to boundary conditions. As an
example consider taking a
periodic lattice, and let the scalar field
be periodic as well.
If the field is
massless, then the zero mode of the field
varies over an infinite range,
in the vacuum state. If we trace over
the degrees of freedom in a
subregion of the lattice, then
the mean value of the field inside is
correlated to the mean value outside, but
this mean value can take on
an infinite range of values. The entanglement
entropy will thus be infinite.

If we take antiperiodic boundary
conditions for the scalar field then the
zero mode does not exist. With this choice (\ref{Ssred})
becomes \cite{SREDPC}
\be S~=~{1\over 3}  \ln(R/a)~+~{1\over 6}\ln (I/R) \ee
where $I$ is the infrared cutoff coming from the finite size of the
lattice.
Another way to kill the zero mode of the scalar field is to have a
vanishing boundary condition for
the field at say $x=0$. Let the traced
over region extend from $x=x_1$ to $x=x_2$. Now modes with
wavelength  much greater than
$x_2$  effectively vanish over
the interval $(x_1,x_2)$, so they do not
serve to entangle this region with
the remainder of the line $x>0$.
Thus the entanglement entropy will be
finite without the need for an
explicit infrared regulator. A third way of
dealing with the zero mode is to
simply assume that the field has a small mass
$\mu$. Then no other infrared
regulation should be needed and the
results should not be sensitive to
choice of boundary conditions.

(ii) For the black hole, we
start by considering a complete
spacelike hypersurface through
the  evaporating geometry, described as follows. Starting at spatial
infinity ($\sigma^-=-\infty$), we move near $\scrip$ to a point with
$\sigma^-=\sigma^-_1< 0 $. (The black hole
vanishes at $\sigma^-= 0$.)
Then we smoothly bend this hypersurface so that it enters the
horizon and reaches the timelike segment of the
line $\Omega=\Omega_{cr}$. (This timelike segment
occurs before this critical line becomes the singularity.) Thus the
hypersurface is kept spacelike all through, and avoids the
singularity by passing below it.
An observer collecting radiation far from the black hole will see the
part of this hypersurface that lies along  $\scrip$, and we would
wish to trace over the part that cannot be observed from outside the
black hole. This would give the reduced density matrix, and thereby
the entropy of entanglement between the field inside and outside the
hole. Let us be more precise about what we take as the 'observed'
part.
Suppose that the observer at $\scrip$
carries an instrument with her which she uses
to collect the outgoing radiation. At first, near $\sigm =-\infty$
nothing
comes out from the black hole. The observer has to wait for quite a
while
before the black hole starts to radiate. At some point the radiation
starts and becomes approximately thermal. We have discussed earlier
that this happens
roughly at $\sigm_0 = -\frac{4\pi M}{\kap \lam^2}$. Correspondingly,
around
this point the observer turns on her instrument.
The observer then collects radiation up to some
retarded time $\sigm_1$, when she turns
her instrument off again.
Thus the part of the hypersurface
between $\sigm_0$ and $\sigm_1$ corresponds to
observations, and the rest of it will be traced over.
Notice that the observer can neither
start nor stop collecting radiation at an
instant, but there will be a short
time scale $d\sigm$ which she needs
to turn on or shut off her instrument in a proper way.
The time interval needed
should be sufficiently rapid to
give a good accuracy for specifying
the turn-on point $\sigm_0$ and the
shut-off point $\sigm_1$, on the other hand
it should not be so small
that it creates disturbances in the matter field
that generate radiation
comparable to the Hawking radiation collected.
It is reasonable to assume that the
time intervals $d\sigm ( \sigm_0)$, $d\sigm ( \sigm_1)$
needed should be given by the typical wavelength
in the outgoing thermal radiation. Thus we can assume
that $d\sigm \sim \beta_H \sim \laminv $ (our
result for the entropy of
the hole will turn out not to depend on this choice).

We split the contribution of different modes to the entropy,
as follows: a) the modes of wavelength $\om^{-1} >> M/\lam^2$:
these modes are effectively constant over the observed region
$(\sigm_0 ,\sigm_1 )$, so they may be taken as a contribution
to the 'zero mode'; b) the leftmoving
modes with $\om^{-1} \lessapp M/\lam^2$;
c) the rightmoving modes with $\om^{-1} \lessapp M/\lam^2$.

Modes of type a) will give a divergence in the entanglement
entropy even in flat Minkowski space (without black hole).
This happens when the mass $\mu$ of the field is taken to
zero, or (if the field is massless) as the observed part of the
hypersurface is taken further and further away from the line
$\Om=\Om_{cr}$ where the field vanishes.
Since we are interested in entropy of entanglement of the
rightmoving Hawking radiation (which occurs over a distance
$M/\lam^2$), we subtract this (diverging) contribution
arising from the large wavelength modes $\om^{-1} >>M/\lam^2$.
We also ignore the leftmoving modes b), as they do not contribute
to the Hawking radiation. The rightmoving modes c) are of
interest to us, but
after particle pairs have been created,  the state of
the field is not the vacuum state for the geometry on the spacelike
hypersurface under consideration.  Srednicki's result, on the other
hand, applies to a vacuum state for the field. The essential idea
is to follow the
rightmoving ({\em i.e.} outgoing)
modes of the field back from $\scrip$, through the collapsing matter
to the line $\Omega=\Omega_{cr}$ where they reflect to left moving
modes
that can be followed back to $\scrim$. Here these left movers are in
the vacuum state, so that we may apply a Srednicki type approach to
estimate the entanglement entropy in these modes.

As we follow  the radiation modes back to $\scrim$ in the manner
indicated
above, we observe the following.
The region $\sigm \in ( \sigm_0 , \sigm_1 )$ in $\scrip$ corresponds
to a region $\yp \in ( \yp ( \sigm_0) , \yp ( \sigm_1 ) )$ in $\scrim$,
where
the relation between $\sigm_i, i=0,1$ and $\yp ( \sigm_i ), i=0,1$
is given by the formula (\ref{eq:rel1}) in section 3.
Thus, the latter region is the region of starting points
for ingoing rays which will
experience redshift and give arise to the
collected radiation in
the 'observed' region of $\scrip$. Also, this region
is separated from the rest of $\scrim$
by 'cuts' of size $d\yp (\sigm_0)$
and $d\yp ( \sigm_1)$, the size of which follows from the size
of the corresponding cuts near $\scrip$ by
the relation (\ref{eq:squeeze1})
given in section 3.
Now we can apply the result of \cite{SRED}.
We disentangle the
finite region
$\yp\in ( \yp ( \sigm_0) , \yp ( \sigm_1 ))$ from
the rest of $\scrim$ by cuts of
size $a_0 =d\yp ( \sigm_0), a_1 =d\yp ( \sigm_1)$.
Ignoring 'zero modes' and the leftmovers,
we expect to create an entropy
for each scalar field (see discussion in section 8 below)
\be
   S = \kap_3 \ln ( \frac{R}{a_0}) +
       \kap_3 \ln ( \frac{R}{a_1}) \ ,
      \label{Sright}
\ee
where $R=\yp ( \sigm_1) - \yp ( \sigm_0)$ and
$\kap_3 = \frac{1}{4}\kap_1$. (A
factor of $\frac{1}{2}$ because we are
considering only rightmovers and
another factor of $\frac{1}{2}$ because
the contribution to $S$ is separated over the two 'cuts'.)

(iii) Let us rewrite (\ref{Sright}) as
\be
 2\kap_3 \ln (R\lam ) + \kap_3 \ln ( \frac{1}{a_0 \lam })
              + \kap_3 \ln ( \frac{1}{a_1\lam }) \ .
\label{Sapprox}
\ee
We substitute $a_i = d\yp( \sigma_i), i=0,1$ given by
the relation (\ref{eq:squeeze1}):
\bea
a_1 &=& \{1 + (e^{4\pi M/\kap \lam} -1)e^{\lam \sigm_1} \}^{-1}
             d\sigm \\ \nonumber
   \mbox{} &\approx & e^{-4\pi M/\kap \lam} e^{-\lam \sigm_1}
                      \laminv \ , \\ \nonumber
a_0 &\sim& \laminv \ .
\eea
The former approximate formula is
valid for $\sigm_1 \in ( -\frac{4\pi M}{\kap \lam^2} , 0 )$ and
the latter just results from the fact that the redshift is negligible
for the rays at earlier times.

After substitution we get (ignoring the infrared cut-off term)
\bea
 S &=& \kap_3 \ln (e^{4\pi M/\kap \lam}e^{\lam \sigm_1})
    +  \kap_3 \ln ( \laminv \lam )
    + 2\kap_3 \ln (R\lam ) \\ \nonumber
   &=& \kap_3 ( \frac{4\pi M}{\kap \lam} + \sigm_1)
    + 2\kap_3 \ln (R\lam )  \ .
\eea
The second term is negligible with respect to the first term, since it
can be calculated that
$R \sim \laminv$ .
The first term is the significant term. As we take $\sigm_1$ closer
and closer to the endpoint $\sigm =0$ (the observer collects more
outgoing radiation), the first term approaches
\be
    \kap_3 \frac{4\pi M}{\kap \lam } \ .
\label{Rslt}
\ee
The above result for entropy must be multiplied by $N$, the number
of scalar fields. Let us deduce
the value of $\kap_3$ by comparing (\ref{Rslt})
with the entropy of entanglement
estimated directly from the
density matrix of the outgoing radiation (eq. (\ref{Stotal})).
This gives
\be
 S \approx \frac{4\pi M}
    {\lam } \ , \  {\rm if} \ \kap_3 =\frac{1}{12}
\label{K3} \ .
\ee
If our assumptions regarding the separation
of the 'zero mode', of right and
left movers are correct ({\em i.e.},
if $\kap_1 =4\kap_3$, $\kap_1$ as given in (\ref{Ssred})),
then (\ref{K3})
agrees with the calculation of
Srednicki which gave $\kap_1 =\frac{1}{3}$.
We discuss this issue further in section 8.

Note that if we consider the evaporation right up to the endpoint, then
the cutoff scale (over which the radiation measurement is switched off)
must go to zero, and the entropy becomes infinite. But since we are
using
the semiclassical geometry, it is not justified to go below distance
$\lam^{-1}$ (or $(N\lam)^{-1}$, for large $N$) in our analysis.
With this restriction, the entropy from the cutoff scale of the endpoint
is ignorable compared to the entropy of the hole, for holes that
evaporate over classical time
scales $4\pi M/ \kap \lam^2 >> \lam^{-1}$.

The significance of the Bekenstein entropy for a black hole is a matter
of debate. One hypothesis is that the horizon behaves as a membrane
with $e^{S_{\rm Bek}}$ states, so that  there is an upper limit  to the
entanglement entropy of the matter outside
the hole with the hole itself \cite{SUSS}. Thus if a
sufficiently large amount of matter were thrown
into the hole then a part of the information would have to leak back out
through subtle correlations in the Hawking radiation \cite{BEK2}.

In a semiclassical treatment of the gravitational field, on the other
hand, there seems to be no limit to the amount of information that can
disappear into the black hole. Thus the entanglement entropy can also
grow without bound when matter is repeatedly thrown into the hole and
the  black hole mass allowed to decrease back to $M$ by
evaporation.\footnote{We are grateful to S. Trivedi for
pointing this out to us.} It is possible to
verify in the simple 1+1 dimensional evaporating
RST solution that the entanglement entropy can indeed exceed the
Bekenstein value by an arbitrary amount.
We illustrate this by taking a simple example: the RST model with two
incoming shock waves, discussed in the previous section.

We again apply the Srednicki approach to estimate entropy.
Consider an observer
at $\scrip$ collecting radiation, who switches on the measuring device
at $\sigm_0$ and switches it off at $\sigm_1$ with the corresponding
switch-on-off intervals $d\sigm (\sigm_i)$ as before. The only
difference
in the two shockwave case is that now the relation between the
$d\sigm $'s at
$\scrip$ and the corresponding $d\yp$'s at $\scrim$ is different. The
relation was given by the formula (\ref{eq:squeeze2})
in section 4. Now we need to substitute
\bea
  a_1 &= & \{1 + (e^{4\pi (M_0+M_1)/\kap \lam} -1)
          e^{\lam \sigm_1} \}^{-1}
              d\sigm \\ \nonumber
   \mbox{} &\approx & e^{-4\pi (M_0+M_1)/\kap \lam}
   e^{-\lam \sigm_1} \laminv \ , \\ \nonumber
 a_0 &\sim & \laminv
\eea
into the equation (\ref{Sapprox}) above. Also
the distance $R$ will be different, but
it is still of the order of $\laminv$ and the $R$-dependent
term can thus be ignored. All this
gives for the entropy (for $N$ fields)
\bea
 S &\approx & N\kap_3 \ln (e^{4\pi (M_0+M_1)/\kap \lam}
     e^{\lam \sigm_1})
    +  N\kap_3 \ln ( \laminv \lam )
    + \ldots \\ \nonumber
   &\approx & N\kap_3 ( \frac{4\pi (M_0+M_1)}
    {\kap \lam} + \sigm_1) + \ldots  \ ,
\eea
where $\mbox{} +\ldots$ represents the
ignored contributions from the
subleading $R$-dependent term
and the infrared cutoff term.
As $\sigm_1 \rightarrow 0$, the entropy becomes
\be
  S \approx N\kap_3 \frac{4\pi (M_0+M_1)}{\kap \lam } \ .
\ee
Substituting $\kap_3 = 1/12$ gives then
\be
  S \approx \frac{4 \pi (M_0+M_1)}{\lam } \ .
\label{Sentangl}
\ee
Note that the Bekenstein entropy, on the other hand, need not
exceed  $S_{Bek}= \frac{2\pi M_0}{\lam }$ at any time if the mass
of the hole never exceeds $M_0$.
The final entropy of
entanglement (\ref{Sentangl}) could be made arbitrarily large (as
one could see by considering {\em eg.} a n-shock wave case).

\bigskip

{\large {\bf 7. Entropy for 3+1 dimensional black holes}}

\bigskip

Let us now turn to the 3+1 dimensional black hole. We will consider
only the Schwarzschild case. The Bekenstein entropy for this hole  is
\be S_{\rm bek}~=~A/4~=~4\pi(2M)^2/4~=~4\pi M^2 \label{SIXONE}
\ee
The entropy collected in the form of
radiation by an observer at $\scrip$ is also
proportional to $M^2$ \cite{Z,PAGE} though it requires use of
the `transmission coefficients' $\Gamma(\om )$ for its computation.

One is tempted to compare such entropies to the result obtained by
carrying out the flat space calculation of Srednicki for the case
of  3 space dimensions. In the latter calculation one considers a
massless
scalar field, say, in the vacuum state in three space dimensions. One
traces over the field modes inside an imaginary sphere of radius $R$,
and computes the entropy of the corresponding reduced density
matrix.
In this calculation it is convenient to decompose the scalar field
into angular modes $Y_{l,m}(\theta,\phi)$, so that we obtain a
1-dimensional problem in the
radial co-ordinate $r$ for each such angular mode. The different
angular modes decouple from each other, so we have to just add the
entropies resulting from the computation for each mode.  The radial
co-ordinate is taken as a 1-dimensional lattice with lattice spacing $a$.
The entropy is  found by numerical computation to be \cite{SRED}
\be S~\approx \ .30 (R/a)^2 \label{SIXTWO} \ee
Thus we seem to reproduce the $\sim R^2$ dependence expected of
the black hole entropy. But if we accept that (\ref{SIXTWO}) applies to
the black hole then we are  faced with the question: What should be
the value of the cutoff $a$?

As we now argue, the result (\ref{SIXTWO}) is not the one relevant
to the 3+1 dimensional black hole.  In fact, the 1-dimensional result
(\ref{Ssred}) is again the relevant one to use.
To see this, consider decomposing the scalar field in the black
hole geometry into modes with angular dependence
$Y_{l,m}(\theta,\phi)$.
Since we have assumed spherical symmetry, these modes decouple
{}from each other. Thus we can consider studying the evolution of
different fields (labelled by $\{l,m\}$) on the 1+1 dimensional geometry
obtained by the spherically symmetric reduction of the 3+1
dimensional geometry.  Proceeding in this way, one would need to
find the `entanglement entropy' of fields in one space dimension, which
(for free fields) is given by the
the result (\ref{Ssred}).

At this point one sees an important difference between the flat
space 3-dimensional problem and the 3+1 dimensional black hole.
This difference occurs in the number of angular modes  $Y_{l,m}$ that
contribute significantly to the entropy. Consider first the flat
space problem. If we had taken a lattice with lattice spacing $a$
all over 3-space, then on the boundary sphere of radius $R$ we could
consider angular modes with
$l \lessapp R/a$. If we first reduce the Hamiltonian to a sum over
modes and then put the $r$ co-ordinate on a lattice, then again it
is found that for $l>>R/a$ the degrees of freedom on the two sides of
the $r=R$ boundary effectively decouple \cite{SRED}. This happens
because the radial  wave equation is
dominated at large $l$ by a  `mass term'  arising from
the angular Laplacian, and such a term
does not couple neighbouring sites of the $r$ co-ordinate lattice.
Since  $-l\le m \le l$, the number of angular
modes contributing to the entropy is $O\{(R/a)^2\}$, which explains
the leading power dependence of the entropy (\ref{SIXTWO}) on the
cutoff $a$.
(Treating the 1-dimensional problem for each angular mode should
give  a $ln(R/a)$ dependence multiplying the dependence $\sim
(R/a)^2$, but the above argument is too crude to distinguish the
presence or
absence of logarithmic terms.)

Thus we see that the difference between
the $a$ dependence of (\ref{Ssred})
and (\ref{SIXTWO})
can be understood in terms of the large number ( $O\{(R/a)^2\}$) of
angular modes contributing to the entropy in the 3-dimensional flat
space problem. But in the 3+1 dimensional black hole, we know that
most of the radiation comes out in only a few angular modes!  In fact
for a
reasonable first estimate of the Hawking radiation one can require
that only the s-wave modes ($l=0$) emerge from the hole. We now
compute the entropy of the 3+1 hole with such an approximation.

While we cannot solve the geometry of the evaporating 3+1
dimensional hole as accurately as for the RST model, for the purposes
of our
calculation we can  consider the Schwarzschild metric with a time
dependent mass $M$.  The surface gravity  of the black hole is
$\kappa=(4M)^{-1}$. The
temperature is $T=\kappa/2\pi=(8\pi M)^{-1}$. The luminosity in the
s-wave mode is
\be L_s~=~{dE\over dt}~=~{1\over 2\pi}\int_0^\infty{ d\om \om\over
e^{\om/T}-1}~=~{\pi T^2\over 12} \label{SIXTHREE} \ee
{}From (\ref{SIXTHREE}) we
compute  $M(y)$, the mass of the hole at  the $\scrip$
Schwartzschild coordinate $y$. (We take $y=0$ at the endpoint of
evaporation, so $y$ is negative in the part of $\scrip$ where the
Hawking radiation is being received.) We have
\be M(y)~=~({-y\over 256 \pi})^{1/3} \ . \ee
As mentioned above, we approximate the evaporating geometry by
just letting $M$ depend on time.  Letting $v$ be the Minkowski
co-ordinate at $\scrim$. This approximation then gives
\be dy~=~-4M(y) d(\ln (v_0-v)) \label{SIX4} \ee
when the co-ordinate $v$ is close to the value $v_0$ which
reflects at $r=0$ to move along the event horizon.
Integrating (\ref{SIX4}) gives
\be \ln (v_0-v_f) -\ln (v_0-v_i)~=~96 \pi (M_f^2-M_i^2) \ee
We have $M_i=M$, and we let $M_f$ be of the order of Planck mass.
Further, $\ln (v_i-v_0)$
can be ignored compared to $\ln (v_0-v_f)$.  Following the
discussion of entropy in the 1-dimensional case, we write
(for one evaporating field, one 'cut' and rightmovers only):
\be S~\approx~-{1\over 12} \ln(a)~=~-{1\over 12} \ln(\delta v)
\label{SIX5}
\ee
Here  $\delta v$ is found from (\ref{SIX4}) by
setting $dy=\Lambda$ (for some
chosen scale $\Lambda$ over which the observation of radiation
is switched off):
\be \delta v ~=~{\Lambda\over 4 M_f}
(v_0-v_f)~\approx~{\Lambda\over 4 M_f}e^{-96\pi M^2} \ee
Substituting in (\ref{SIX5}) we get
\be S~\approx~ 8\pi M^2~=~2 S_{\rm Bek}
\label{SIX6}
\ee
We have ignored terms logarithmic in $M$; these corrections are
smaller than the contribution of the $l \ne 0$ modes which we have
also neglected.

We can now compare the result (\ref{SIX6}) for the
entanglement entropy with
the thermodynamic entropy collected at $\scrip$.  Following \cite{Z},
we  compare the change in the  thermodynamic entropy
received at  $\scrip$ to the change
in the Bekenstein entropy of the hole:
\be R~=~dS/dS_{\rm Bek}~=~{\int_0^\infty dx x^2
\sigma(x)\{xe^x/(e^x-1)-\ln(e^x-1)\}\over \int_0^\infty dx x^3
\sigma(x)/(e^x-1)}
\label{SIX7}
\ee
Here
\be \sigma(\om)~=~\sum_{l,m} \Gamma_{l,m}(\om)/[27(\om M)^2]\ee
is the absorptivity per unit area of the black hole. In the above
used approximation to Hawking radiation we used only the $l=0$
mode, and let the `transmission coefficient'
$\Gamma$ be unity for all  $\om$.
This gives $\sigma(\om)=1/[27(\om M)^2]$. Substituting
this in (\ref{SIX7})
gives $R=2$, in accordance with (\ref{SIX6}).
For more realistic models, taking into a account the
transmission coefficients $\Gamma_{l,m}(\om)$, one obtains
$R\sim 1.3-1.6$ \cite{PAGE2}. To reproduce the effect of nontrivial
$\Gamma (\om)$ we would need to extend
the Srednicki calculation to fields
with position dependent 'mass term'; we do not
investigate this further here.

We can also compare the above derivations  to the direct
computation of the
entropy of the density matrix obtained in the evaporation process;
{\em i.e.}, carry out the calculation
analogous to (\ref{roo}) to (\ref{Stotal}).  We again
have
$S_n=\pi^3T/3\epsilon$, but now $T=(8\pi M_n)^{-1}\equiv T_n$.
Following the evaporation process we find
$M_n=(n/128\epsilon)^{1/3}$. This gives, as expected,
\be S~=~\sum_{n=1}^{n_{\rm max}}
{\pi^3 T_n\over 3\epsilon}~\approx ~8\pi M^2~=~2S_{\rm Bek} \ .\ee

In our above application of Srednicki's result, we found that the
one space dimensional formulation was more relevant, rather than
the three space dimensional one. On the other hand if an observer
stands near the horizon of the black hole, she sees thermal radiation
with power in a large number of angular momentum modes. Then it is
possible that by carrying out the above calculations with a different
choice of hypersurface ({\em e.g.} with the 'outside' part of the
hypersurface corresponding to a static frame near the horizon) one
would find relevant the equation (\ref{SIXTWO}).

\bigskip

{\large {\bf 8. Discussion}}

\bigskip

In this paper we have carried out the computation of Bogoliubov
coefficients, stress tensor and the entanglement entropy for an
evaporating black hole. Before one had explicit models of evaporating
geometries, such quantities had been calculated only in the absence
of backreaction. With backreaction, it is possible to obtain
for example the stress tensor in the region between the event
horizon and the apparent horizon.

Concerning the application of Srednicki's result to
the black hole entanglement entropy, we have made two assumptions.
First, we have assumed that the infrared divergence
comes from modes with wavelength
very large compared to the system size; after these modes are
removed, the entropy can be split into a contribution from
rightmoving modes and a contribution from leftmoving modes.
Second, we assumed that when we cut the region R out of a line,
the entropy of system, say, can be split into two parts, one
coming from each of the two cuts at the two ends of R.
What we do now is offer some heuristic arguments to justify these
assumptions.

First we wish to understand the appearance of the logarithmic
dependence in the entanglement entropy. Consider a segment of the
real
line, $0 \leq x \leq I_1+I_2$, divided into two regions
near $x=I_1$ by a 'cut' of length $a <<I_1,I_2$. Further,
assume $I_1<I_2$. The scalar field we take to vanish at
$x=0$, $x=I_1+I_2$.  What is the entanglement entropy of
$I_1$ with $I_2$ (with cutoff scale $a$)?

Suppose $I_1=I_2$. The field modes have wave numbers
\be
   k = \frac{n\pi}{I_1+I_2} \ , \ n=1,2,\ldots \ .
\label{Eight1}
\ee
The mode $k =\pi /(I_1+I_2)$ is 'shared' between the
two sides $I_1$,$I_2$, and we assume that it gives an
entropy $s$. Now consider modes with
$1 \lessapp n \lessapp (I_1+I_2)/a$. For any scale
$\om^{-1}$ of the wavelength we can make localized
wavepackets just as was done in section 6
(eq. (\ref{packet})). The number of
such wavepackets is $\sim n$. Only the wavepacket that
overlaps both sides of the cut at $x=I_1$
contributes to the entanglement entropy; and
we again take this entropy to be $s$.
(This is an assumption.) Equivalently,
we find that each mode (\ref{Eight1})
contributes $\frac{s}{n}$ to the entropy,
which becomes
\be
  S \approx s \sum^{I_1/a}_{n=1} \frac{1}{n}
    \approx s \ln \frac{I_1}{a} \ .
\label{Eight2}
\ee
Now suppose instead that $I_1 <<I_2$.
For $k^{-1} >>I_1$, the mode essentially
vanishes over $I_1$, and so cannot entangle
this region with $I_2$. For
wavelengths $k^{-1} \lessapp I_1$
we make localized wavepackets in the region
$0<x<\alpha x$, $\alpha \greapp 1$, just as in the
discussion above, and
thus find again eq. (\ref{Eight2}), where
we note that $I_1$ is the smaller segment.

Now we address a more complicated case. We have
the segments
\begin{itemize}
\item $I_1:\ 0<x<I_1$
\item $R:\ I_1<x<I_1+R$
\item $I_2:\ I_1+R<x<I_1+I_2+R \ . $
\end{itemize}
Let $R<<I_1<<I_2$. $I_1$ and $R$ are separated by a cut
of size $a_1<<R$, and $R$ and $I_2$ are separated by
$a_2<<R$. We want the entanglement entropy of $R$ with the
remainder
({\em i.e.}, $I_1 \cup I_2$).
Again, we do not need to consider modes with wavelength
$k^{-1} >>I_1$, since these effectively vanish over $R,I_1$.
The modes with $R <k^{-1} <I_1$ give a contribution
$S \approx s \ln \frac{I_1}{R}$. The modes with $k^{-1} \lessapp R$
can be made into wavepackets that get 'partitioned' at two places;
$x=I_1,x=I_1+R$. These give the
contributions $S \approx s\ln \frac{R}{a_1}$,
$S \approx s\ln \frac{R}{a_2}$. Overall, we then get
\be
 S \approx s\ln \frac{R}{a_1} + s\ln \frac{R}{a_2} +
     s\ln \frac{I_1}{R} \ .
\ee
If $a_1=a_2\equiv a$, $I_1=I_2\equiv I$, we can write
\be
 S \approx 2s\ln \frac{R}{a} + s\ln \frac{I}{R} \ ,
\ee
which resembles (\ref{Ssred}) with $\kap_2 =-\frac{1}{2}\kap_1$.
The value of $s$ we can fix by the direct calculation
of the entropy in the 1+1 dimensional
black hole evaporation preceeding eq. (\ref{Stotal}).
For the modes with $k^{-1} \lessapp R$, one can clearly make
a breakup between right and left movers. Thus we conclude
$s=\frac{1}{6}$,
after doubling up the obtained answer for the rightmovers alone.

The above gives a heuristic understanding of the entropy of
entanglement,
which should be useful in applying the result with a variety of boundary
conditions.

In conclusion, we note that the 'exponential expansion' of coordinates
near the horizon gives rise to thermal radiation, as was shown by
Hawking. By using the result of Srednicki, the same coordinate
transformation
gives the entanglement entropy produced by the thermal radiation.
Thus we seem to be one step closer to understanding the nature of
black hole thermodynamics.

\bigskip

{\large {\bf Acknowledgements}}

\bigskip

We thank M. Crescimanno, A. Dabholkar, C. Itzykson and G. Lifschytz
for many
discussions. We are grateful to S. Trivedi for discussions and
for informing us of the reference \cite{Z}.

This work was supported in part by funds provided by the U.S.
Department of Energy (D.O.E.) under
contract $\sharp$DE-AC02-76ER03069.

\newpage

{\large {\bf Figure caption}}

\bigskip

{\bf Figure 1.} The black hole geometry formed by an incoming shock
wave (thick line with arrows).
Evaporation occurs in region II; regions I and III are
linear dilaton vacuua.

\end{document}